\documentclass[letterpaper,10pt]{article}%draftcls,peerreview]{IEEEtran}%{article}%

\usepackage{graphicx}
\usepackage{amsmath}
\usepackage{amssymb}
\usepackage{cite}

\newcommand {\F}{\mathbb{F}}
\newcommand {\G}{\mathcal{G}}
\newcommand {\ie} {\textit{i.e.}}
\newcommand {\trace}{\textrm{Trace}}
\newcommand {\conj}[1] {\overline{#1}}
\newcommand {{\Chi}} {{\chi}}
\newcommand{\R}{\mathbb{R}}
\newcommand{\C}{\mathbb{C}}
\newcommand{\Code}{\mathcal{C}}
\newcommand{\Cl}[2]{ \textrm{Cl}_{#1}({#2})}

\begin{document}
%---------
% Counter
%---------
\newcounter{tcount}
\setcounter{tcount}{0}

%----------------------------
%Personal new environements
%----------------------------

\newtheorem{thm}{Theorem}
\newtheorem{coro}[thm]{Corollary}
\newtheorem{lemme}[thm]{Lemma}
\newtheorem{nota}[thm]{Notations}
\newtheorem{defi}[thm]{Definition}
\newtheorem{prop}[thm]{Proposition}
\newtheorem{rmq}[thm]{Remarks}

\newenvironment {proof} {\noindent{\bf Proof :}} {} {}
\newenvironment {keywords} {\noindent{\bf Keywords :}} {} {}

%--------------------
% Title informations
%--------------------
\title{Constructions of Grassmannian Simplices}
\author{Jean~Creignou}% <-this % stops a space
\maketitle

%--------------
% End of TITLE
%--------------

%----------------------
% ABSTRACT & KEY WORDS
%----------------------
\begin{abstract}

 The aim of this article is to present new and explicit constructions
 of optimal packings in the Grassmannian space. Therefore we use a
 method presented in \cite{AGroup} involving finite group representations.
 Infinite families of configurations having only one non-trivial set of
 principal angles are naturally found using
 2-transitive groups. These packings are proved to reach the simplex
 bound \cite{SloaneArt} and are therefore optimal w.r.t. the chordal
 distance. The construction is illustrated by several examples.
\end{abstract}


%         -------------------
%         -------------------
%              THE BODY 
%         -------------------
%         -------------------

%-----------------------
% PART I Introduction
%-----------------------

\section{Introduction}
% The problem of finding `good' configurations of
% $m$-di\-men\-sion\-al subspaces in $\R^n$ or $\C^n$ has 
% various applications. For example in the field of Information
% Theory dealing with (non-coherent) multiple-antennas transmissions where the use of
% multiple antennas and suitable Grassmannian codes enables diversity and multiplexing gain
% \cite{TSE,MR1607687,MR1873925,MR1748987,TSE_ART,MR2245117}.\\

 Space time codes have been fiercely studied since the publication of 
 \cite{Marzetta} and \cite{Telatar} where the ability of such codes 
 to enhance communications is revealed. Part of this interest has turned over
 Grassmannian coding when the link between the non-coherent case and Grassmannians
 was clearly explained by Zheng and Tse in \cite{TSE_ART}.
 Several papers (as \cite{MR2245117}) have shown that if the so-called product distance is linked to
 the performance of the code at high SNR (Signal to Noise Ratio), the chordal distance
 is a key parameter for performance at low SNR . 
 Since the codes described in this article are optimal regarding the chordal distance 
 as they meet the simplex bound introduced in \cite{SloaneArt}, one may 
 expect them to enhance communications especially when the channel is very noisy.
 The notion of chordal distance has first appeared in \cite{SloaneArt} where
 general constructions for Grassmannian packings are described. We were particularly 
 interested in the following classical construction using group orbits.
 Let $W$ be a $m$ dimensional subspace in $\C^n$ and let $G$ be a finite subgroup of $U_n(\C)$
 then we can consider the Grassmannian code made of the orbit of $W$ under $G$.

 Constructions using group orbits have motivated coding theorists for years
 and have already been successfully used in the context of spherical codes or codes
 over finite fields. In the Grassmannian case as others the result of 
 such a construction strongly depends on the group and 
 the first element of the orbit. This choice arises then as the main problem.\\

 There are two kind of codes obtained as the orbit of an element 
 under a group $G$. Codes having a cardinality equal to 
 the cardinality of the group and codes where the stabilizer of an element
 is a non-trivial subgroup of $G$. This second kind of codes are naturally
  more structured and smaller. The originality of this article is to take
 the problem in a unusual way. We first search in 
 the group $G$ which subgroups are interesting 
 candidates for a stabilizer then we deduce a convenient starting element.\\

 To compare Grassmannian codes obtained by various ways, we introduce the parameters
 of a Grassmannian configurations as the numbers $[n,m,N,d]$, where $m$ is the dimension
 of the subspaces, $n$ the dimension of the surrounding space, $N$ the number of elements,
 and $d$ the minimal (chordal) distance. In this article we compute the parameters of Grassmannian
 packings obtained by the use of 2-transitive groups and subspaces coming from representation
 theory and show that they meet the so-called simplex bound (Thm \ref{main_thm}) providing optimal packings.

 We first recall in Section 2 basic facts concerning Grassmannian
 spaces and representation theory. Then we present how to use 2-transitive groups 
 and representation theory to construct
 special configurations which reach the simplex bound. Section 3 gives
 a detailed formulation of this result and its proof. Explicit examples coming from the classification 
 of 2-transitive groups are developed in Section 4. We end with some generalizations in Section 5 and  conclude.\\

\section{Basic facts on Grassmannian spaces and representation theory}
%---------------------------------------
% SUBPart 1  (Grassmannian Introduction)
%---------------------------------------
\subsection{Grassmannian spaces}\label{GRASS}

 In this section we recall from \cite{SloaneArt}\footnote{Warning : their 
 notations for $n$ and $m$ are different from ours.} the
 main background about Grassmannian spaces used in the
 sequel. We introduce Grassmannian spaces, define distances
 between elements and recall the expression of the simplex
 bound.\\

 The Grassmannian space over the complex numbers denoted by $\G_{m,n}$ is simply the set of
 all $m$-dimensional vector subspaces of $\C^n$. To introduce a distance
 between subspaces the notion of principal angles is needed.\\

\begin{defi}
 For any $\mathcal{P,Q}$ elements of $\G_{m,n}$ (\ie{} two 
 $m$-dimensional subspaces  in $\C^n$), let\footnote{ By ``arg max" we mean
 the arguments (any) which allow the following function to reach it's
 maximum.}
 $$x_1,y_1:=\textrm{arg} \max_{\substack{x\in \mathcal{P},y\in\mathcal{Q}\\||x||=||y||=1}} |\langle x,y\rangle|.$$
 Then by induction, constrain $x_i$ (resp. $y_i$) to be a unit vector
 orthogonal to  $\{x_1,\dots,x_{i-1}\}$ (resp. $\{y_1,\dots,y_{i-1}\}$) and
 define
 $$x_i,y_i:=\textrm{arg} \max_{\substack{x\in \mathcal{P},y\in\mathcal{Q}\\ \textrm{constrained}}} |\langle x,y\rangle|,\quad i \le m.$$
 The principal angles are the values 
 $\theta_i \in [0,\frac{\pi}2]$  such that
 \mbox{$\cos(\theta_i)=|\langle x_i,y_i \rangle|$}, $i=1,...,m$.\\
\end{defi}

 \noindent One can prove that the set of principal angles $(\theta_1,...,\theta_m)$ characterizes 
 the orbit of the pair $(\mathcal{P,Q})$ under the action of the orthogonal group. 
 To define distance between pairs we consider the following definitions :

  $$\begin{array}{rl}
\displaystyle d_c(\mathcal{P,Q})&:= \displaystyle \sqrt{\sum_i \sin^2(\theta_i)} \\
\\
\displaystyle \tilde{d}(\mathcal{P,Q})&:= \displaystyle \prod_i \sin(\theta_i).\\
\end{array}$$

 \noindent In this paper we mainly focus on the chordal distance $d_c$ but 
 we also refer to $\tilde{d}$. It is not a distance 
 in the mathematical sense, but it is the key criterion  
 for estimating the performance of a wireless communication at high SNR
 \cite{MR2245117}. 

 The chordal distance can also be expressed in an easy way using trace and
 projection matrices \cite{SloaneArt}. Let $\Pi_\mathcal{P}$ and
 $\Pi_\mathcal{Q}$ be the projection matrices on $\mathcal{P}$ and
 $\mathcal{Q}$ then,
\begin{equation}\label{chordaltrace}
 d_c^2=\frac{1}{2}|| \Pi_\mathcal{P}-\Pi_\mathcal{Q}||_2^2=\textrm{
 Trace}(\Pi_\mathcal{P})-\textrm{Trace}(\Pi_\mathcal{P}\Pi_\mathcal{Q}).
\end{equation}

 The chordal distance has another advantage. As shown in \cite{SloaneArt}, 
 this metric enables an embedding of the Grassmannian space in a sphere
 of higher dimension. One can then deduce bounds for codes
 in Grassmannian spaces from bounds for spherical codes. We
 recall here the \emph{simplex bound} on Grassmannian configuration
 (obtained by this very way). This bound was stated for real
 Grassmannian spaces but extends itself in an easy way to
 the complex case\footnote{Indeed any complex configuration
 can be embedded into a real space doubling $m$, $n$ and
 $d_c^2$. After simplifications one obtains the same bound.}.\\

 \begin{lemme}\label{lemme_simplex_bound} [Simplex Bound] \cite{SloaneArt} 
  For any configuration of $N$ subspaces of dimension $m$ in $\C^n$, the
  following inequality holds :
  \begin{equation}\label{RankinSimplexBound}
  d_c^2 \le \frac{m(n-m)}{n}\frac{N}{N-1}\end{equation}
  $$\textrm{ equality  requiring } N\le \binom{n+1}{2}.$$
 \end{lemme}

 If equality in (\ref{RankinSimplexBound}) occurs then the distance between 
 each pair of distinct elements is the same  \cite{SloaneArt}. As an extension
 from the spherical case the term \emph{simplicial} is used for such a configuration.
 One has to be careful since it is not true that any \emph{simplicial} configuration
 meet the simplex bound (\ref{RankinSimplexBound}).
 This paper deals with a particular case of simplicial configurations 
 having only one non-trivial set of principal
 angles between any pairs of $m$-dimensional planes. 
 We use the term \emph{strongly simplicial} to denote configurations having
 only one non-trivial set of principal angles\footnote{Similarly such configurations 
 do not automatically reach the simplex bound.}.\\

 Next section recalls results of representation theory used in the sequel and
 sets some notations.
 
%---------------------------------
% SUBPart II Representation theory
%---------------------------------
 
 \subsection{Representation theory}\label{REP_THEO}
 The constructions we want to study are based on representation theory.
 Some good references on the subject are
 \cite{MR1153249,MR1280461,MR1864147,MR793377,MR543841,MR1695775}
 where one can find more details. We give here a  brief summary of results
 and definitions.\\

\begin{defi}
 A unitary (complex) matrix representation of a finite group $G$ is a morphism,
 $\rho : G \to U_n(\C)$. The dimension of the matrices is called the
 dimension or the degree of the representation. The function 
 $\Chi_\rho : G \to \C$ defined by $\Chi_\rho(g)=\trace(\rho(g))$ is the
 character associated to $\rho$. It is worth noticing that $\Chi_\rho(1)$
 is equal to the dimension of $\rho$ and that $\Chi_\rho$ is constant on
 the conjugacy classes of $G$. \\
\end{defi}

 Two representations $\rho,\ \rho'$ are said to be equivalent if and only if
 there exists an invertible matrix $U$  such that
 $\forall g\in G,\ U.\rho(g).U^{-1}=\rho'(g)$. As a consequence, two equivalent
 representations have equal characters. Moreover, one can show that conversely
 two representations having the same character are equivalent.

 A representation $\rho$ is called reducible if there exists a proper subspace $W$
 $(0\subsetneq W \subsetneq \C^n)$ such that 
 $\forall g \in G,\ \rho(g)W\subset W$. It is irreducible otherwise, this vocabulary extends to 
 characters\footnote{The character of an irreducible representation is called an irreducible character.}.
 When a representation $\rho$  is reducible, the $\C$-vector space $\C^n$ can be decomposed as a direct sum of 
 orthogonal subspaces
 isomorphic to irreducible representations (Maschke Theorem \cite{MR1280461,MR1864147}).
 This decomposition is not unique but the number of subspaces isomorphic to a given irreducible representation
 $\rho_0$ is well defined and called the multiplicity of $\rho_0$ in $\rho$. 
 The direct sum of all subspaces isomorphic to a given irreducible
 representation is called an isotypic subspace and moreover the decomposition of  $\C^n$ 
 in isotypic subspaces is unique \cite{MR1280461,MR1864147}. An isotypic subspace $W$ is naturally 
 characterized by an irreducible character $\Chi$ of $G$ and has this projection matrix \cite{MR1280461,MR1864147}:
\begin{equation}
 \Pi_W=\frac{\Chi(1)}{|G|}\sum_{g \in G}  \conj{\Chi(g)}\rho(g).
\end{equation}

 There are as many (non-equivalent) irreducible characters of $G$ as conjugacy  classes in $G$.
 The set of irreducible characters $\{\Chi_i : i=1,\dots,t\}$, have the following properties 
(the so-called orthogonality relations)  :
 
 $$
 \frac{1}{|G|}\sum_{g\in G} \Chi_i(g)\conj{\Chi_j(g)} =
 \left\{\begin{array}{l} 1 \textrm{ if } i=j \\
 0 \textrm{ if } i \ne j \end{array}\right. ,
 $$

 \begin{equation}\label{orthorelations}
 \frac{1}{|G|}\sum_{i=1}^t \Chi_i(g)\conj{\Chi_i(h)} =
 \left\{\begin{array}{l} |\Cl{G}{g}| \textrm{ if } \Cl{G}{g}=\Cl{G}{h}\\
 0 \textrm{ if } \Cl{G}{g}\ne \Cl{G}{h}
 \end{array}\right.,
  \end{equation}
 where  $\Cl{G}{g}$ denotes the conjugacy class of $g$ in $G$, and $|.|$ is a notation for cardinality.

 As a direct consequence, the set of irreducible characters forms an orthogonal basis 
 of the space of class-functions\footnote{A class
 function is a function which is constant on conjugacy classes.} regarding
 the inner product :
 $$\langle\Chi,\psi\rangle=\frac{1}{|G|} \sum_{g\in G} \Chi(g)\conj{\psi(g)}.$$

 The restriction of an irreducible character $\chi_\rho$ of $G$ to a subgroup $H\subset G$ 
 is noted $\chi_\rho\downarrow_H^G$. This new character may now be reducible and the representation
 $\rho\downarrow^G_H$ contains irreducible
 representations with multiplicities. If these multiplicities are denoted $\lambda_i$ we may write
 
 $$\chi\downarrow_H^G=\sum \lambda_i \chi_i$$

 In order to avoid confusion in the sequel we prefix words with the group we want refer to
 (we use for example the term ``$H$-irreducible"  for representations irreducible w.r.t the group $H$,
 ``$H$-isotypic" for isotypic subspaces w.r.t. the group $H$,...).\\

 To lighten expressions when the context allows no misleading we note $gW$ instead 
 of $\rho(g)W$ for the action of $\rho(g)$ on the subspace $W$.\\

%---------------------------------------------
% PART IV Description and proof of optimality
%---------------------------------------------
 
 \section{Code construction and optimality}\label{labelresults}

 \qquad The main result of this paper is the construction of \emph{strongly simplicial}
 packings in Grassmannian
 spaces which reach the simplex bound and are therefore optimal w.r.t. the
 chordal distance. In this section we describe the construction and its elementary properties then 
 we prove that 2-transitive groups acting on direct sums of isotypic subspaces give optimal 
 configurations regarding the chordal distance.\\

%-----------------------------------------------
% SUBPART General construction and  properties
%-----------------------------------------------

% \subsection{General construction and  properties}

 The optimal packings we want to discuss come from the following construction
 given in \cite{SloaneArt}.\\
 
 \begin{defi}\label{CONSTRUCTION0}
 Let $\rho\ :\ G \to U_n(\C)$ be an irreducible representation and $H$ a subgroup of $G$.
 Take a non trivial subspace $W \subset \C^n$ of dimension $m$ which is stable under the action of $\rho(H)$.
 We denote by $\Code(W,G)$ the orbit of $W$ under the action of $G$. The cardinality of this code is
 $|G/H'|$, where $H'$ is the stabilizer of $W$ in $G$.\\
 \end{defi}

 Without loss of generality we can assume that $H$ is the stabilizer of $W$ in $G$.
 Then the code $\Code(W,G)$ has the nice following property :\\

 \begin{prop}\label{EQANGLES}
  The number of different sets of principal angles in a code $\Code(W,G)$ is
  bounded above by the number of orbits for the action of  $G$ on unordered pairs of $G/H$. Moreover the set of principal angles 
  for the pairs $(W,g_0W)$ and $(W,gW)$  is the same for all $g\in Hg_0H \sqcup  Hg_0^{-1}H$. \\
 \end{prop}
 \begin{proof}
 Let $\textrm{SPA}(W_1,W_2)$ be the set of principal angles between two subspaces $W_1$ and $W_2$.
 Since a unitary transformation preserves the set of principal angles (see Section \ref{GRASS}) it is easy 
 to see that for any $g,g_1,g_2$ in $G$ and any $h_1,h_2$ in $H$, 
 $$\textrm{SPA}(g_1W,g_2W)\ \ = \ \ \textrm{SPA}(gg_1h_1W,gg_2h_2W)\ \ =\ \ 
 \textrm{SPA}(W,h_1^{-1}g_1^{-1}g_2h_2W).$$
 The proposition follows directly.\\
 \end{proof}

 After this proposition it is natural to look at groups $G$ and subgroups $H$ such that there are few double
 classes $H\backslash G /H$. The codes obtained with such groups have few distinct values for the distances between subspaces.
 This is true for any `distance' definition which depends on the set of principal angles in particular for $d_c$ 
 and $\tilde{d}$.

 \begin{coro}
   If $G$ is 2-transitive on $G/H$ (or otherwise stated : if $|H\backslash G /H|=2$)
   then there is only one non-trivial set of principal angles.\\
 \end{coro}

%---------------------------------------------------
% SUBPART Optomality regarding the chordal distance
%---------------------------------------------------

% \subsection{Optimality regarding the chordal distance}

 Using group orbits to construct codes is a standard method. In the Grassmannian case as others the result depends strongly on the choices of starting element(s) and groups. In this context the choice of isotypic subspaces and 2-transitive groups is quite interesting as shows the following theorem.

 \quad \begin{thm}\label{main_thm}
 Let $G$ be a group and $H$ a subgroup such that $G$ acts 2-transitively by left
 multiplication on $G/H$. Suppose furthermore that we dispose of an
 irreducible representation  
 $\rho : G \to \textrm{GL}_n(\C)$ which is reducible when restricted to
 $H$.
 Under these hypothesis let $W$ be a direct sum of $H$-isotopic
 subspaces\footnote{ By $H$-isotypic subspaces we mean isotypic 
 subspaces relative to the restriction of $\rho$ to the subgroup $H$.},
 then the orbit of $W$ under $G$ forms a 
 \emph{strongly simplicial}\footnote{We recall that by this we mean that 
 there is only one non-trivial set of principal angles between any pair of
 distinct elements.} Grassmannian configuration reaching the simplex bound.
 The parameters are $N=|G/H|$ ; $n=\dim(\rho)$ and $m=\textrm{dim}(W)$.\\
\end{thm}

%---------------------------
% SUBPART Proof of main thm
%---------------------------

 %\subsection{Proof of main theorem}
 If $G$  is 2-transitive on $G/H$ then $H$ is a maximal subgroup of $G$.
 So the cardinality of the code $\Code(W,G)$ 
 is equal to $|G/H|$. The only result left to prove is that if $W$ is a direct sum of isotypic subspaces
 then the simplex bound is reached. For this we  need three
 lemmas. The first gives another expression of  the value of the chordal
 distance in our constructions. The two others deal with sums related to 
 character  theory.\\

 \begin{lemme}
 Let $W$ be the direct sum of isotypic subspaces, associated to a subset of
 H-irreducible characters $\{\Chi_1,...,\Chi_s\}$. Then the square value
 of the chordal distance between $W$ and $gW$  is 
 \begin{equation}\label{fonda2}
 d_c^2=  \displaystyle \sum_{i=1}^s \lambda_i {\displaystyle
 \Chi}_i(1) \displaystyle -\:\frac{1}{|H|^2}\sum_{h_1,h_2\in H}
 E(h_1)E(h_2){\displaystyle \Chi}_{\rho}(h_1gh_2g^{-1})
\end{equation} where
 $$
 E(h):=\left(\sum_{i=1}^s{\displaystyle \Chi}_i(1)\conj{{\displaystyle  \Chi}_i(h)}\right)
 $$
 and $\lambda_i$ is the multiplicity of $\Chi_i$ in the restriction of $\Chi_\rho$ to $H$ (see Section \ref{REP_THEO}) :
 $$\Chi_\rho\downarrow^G_H= \sum_i \lambda_i \Chi_i.$$
 \end{lemme}

\begin{proof}
 The projection matrices on $W$ and on $gW$ are
 $$\Pi_W=\sum_{i=1}^s\frac{\Chi_i(1)}{|H|}\sum_{h \in H}
 \conj{\Chi_i(h)}\rho(h),$$
 $$\Pi_{gW}= \rho(g) (\Pi_W) \rho(g^{-1}) =\sum_{i=1}^s \frac{\Chi_i(1)}{|H|}\sum_{h \in H}
  \conj{\Chi_i(h)}\rho(ghg^{-1}).$$
 The result follows from these expressions, formula (\ref{chordaltrace}) for
 chordal distance and the orthogonality relations (\ref{orthorelations}).
 Indeed we have
 $$\trace(\Pi_W)=\sum_{i=1}^s \lambda_i\Chi_i(1)$$ 
 and
  $$\trace(\Pi_W\Pi_{gW})=\frac{1}{|H|^2} \sum_{h_1,h_2\in H} E(h_1) E(h_2) \trace(\rho(h_1)\rho(gh_2g^{-1})).$$ 

\end{proof}

 We now give two character formulas. We also need to introduce further
 notations.\\
 
\begin{nota} For any set $S$, the formal sum of all elements in $S$ is
 written $\widehat{S}$. For a character $\Chi(\widehat{S})$ means 
 $\sum_{s \in S} \Chi(s)$. As a consequence if $\Chi$ is a character of
 $G$ then $\Chi(\widehat{\Cl{G}{h_1}})=|\Cl{G}{h_1}| \Chi(h_1)$.\\
\end{nota}

\begin{lemme} For any irreducible character $\Chi_{\rho}$ on $G$,
\begin{equation}\label{XiXi}
\Chi_{\rho}\left(\widehat{\Cl{G}{h_1}}\widehat{\Cl{G}{h_2}}\right)=\frac{|\Cl{G}{h_1}||\Cl{G}{h_2}|\Chi_{\rho}(h_1)\Chi_{\rho}(h_2)}{
\Chi_{\rho}(1)}.
\end{equation}
\end{lemme}

\begin{proof} The following relation can be found in \cite{MR1864147},
 chapter 30. If $\textrm{Cl}_1,...,\textrm{Cl}_\ell$ are all conjugacy classes in $G$ then
$$ \widehat{\textrm{Cl}}_i\widehat{\textrm{Cl}}_j=\sum_{k=1}^{\ell} a_{ijk} \widehat{\textrm{Cl}}_k$$
where
 $$ a_{ijk}=\frac{|\textrm{Cl}_i||{\textrm{Cl}}_j| }{|G|} \sum_{\Chi \textrm{ irred.}} \frac{\Chi(\textrm{Cl}_i)\Chi(\textrm{Cl}_j)\conj{\Chi(\textrm{Cl}_k)}}{\Chi(1)}.$$
% So
%$$\Chi_{\rho}(\widehat{\Cl{G}{h_1}} \widehat{\Cl{G}{h_2}})  =   
%\displaystyle \sum_{k=1}^\ell
%\frac{|\Cl{G}{h_1}| |\Cl{G}{h_2}}{|G|}  \sum_{\Chi \textrm{ irred.}}
%\frac{\Chi(\Cl{G}{h_1})\Chi(\Cl{G}{h_2})\conj{\Chi(\textrm{Cl}_k)}}{\Chi(1)}
%\Chi_{\rho}(\widehat{\textrm{Cl}}_k)$$

%$$\Chi_{\rho}(\widehat{\Cl{G}{h_1}}\widehat{\Cl{G}{h_2}})= 
%\displaystyle 
% \sum_{\Chi \textrm{ irred.}} \frac{|\Cl{G}{h_1}| |\Cl{G}{h_2}}{|G|}
%\frac{\Chi(h_1)\Chi(h_2)}{\Chi(1)} \displaystyle 
%{\sum_{k=1}^\ell \conj{\Chi(\textrm{Cl}_k)}
%\Chi_{\rho}(\widehat{\textrm{Cl}}_k)}.$$
%Since  $\sum_{k=1}^\ell \conj{\Chi(\textrm{Cl}_k)}
%\Chi_{\rho}(\widehat{\textrm{Cl}}_k)=\left\{
%\begin{array}{l} 0 \quad
%\textrm{ if } \Chi\ne \Chi_{\rho}\\
% |G| \quad \textrm{ if } \Chi=\Chi_{\rho}\end{array}\right.$ from first 
% orthogonality relation (\ref{orthorelations}), the formula is proved. 

%%% MODIF
Formula (\ref{XiXi}) follows immediately from this result and orthogonality relations (\ref{orthorelations}).
%%%

\end{proof}

\begin{lemme} \label{formule3}
For any irreducible character $\Chi_{\rho}$ on $G$,
 \begin{equation}\label{Ancien_A}
 \sum_{g\in G}
 \Chi_{\rho}\left(h_1gh_2g^{-1}\right)=\frac{|G|\Chi_{\rho}(h_1)
 \Chi_{\rho}(h_2)}{\Chi_{\rho}(1)}.
 \end{equation}\\
\end{lemme}

\begin{proof}
\begin{equation}\label{valeurA1}
\sum_{g\in G}
 \Chi_{\rho}\left(h_1gh_2g^{-1}\right)=\frac{|G|}{|\Cl{G}{h_2}|}\Chi_{\rho}(h_1 \widehat{\Cl{G}{h_2}})
\end{equation}

\noindent but also 
\begin{equation}\label{valeurA2} \sum_{g\in G}
 \Chi_{\rho}\left(g^{-1} h_1gh_2\right)=\frac{|G|}{|\Cl{G}{h_1}|}\Chi_{\rho}(\widehat{\Cl{G}{h_1}}h_2)
\end{equation}

\noindent From (\ref{valeurA1}) and (\ref{valeurA2}) one can see that 
 $\sum_{g\in G} \Chi_{\rho}\left(h_1gh_2g^{-1}\right)$ does not depend of
 $h_1 \in \Cl{G}{h_1}$ or $h_2 \in \Cl{G}{h_2}$, so 

\begin{equation}\label{valeurA3}
 \sum_{g\in G}
 \Chi_{\rho}\left(h_1gh_2g^{-1}\right)=\frac{|G|}{|\Cl{G}{h_1}||\Cl{G}{h_2}|}\Chi_{\rho}(\widehat{\Cl{G}{h_1}}\widehat{\Cl{G}{h_2}}).
\end{equation}
\noindent Equation (\ref{valeurA3}) together with (\ref{XiXi}) gives the result.\\
\end{proof}

We are now ready to prove Theorem \ref{main_thm}.\\

\begin{proof} [Thm \ref{main_thm}]
 Without loss of generality we assume that $W$ is the isotypic subspace
 associated to $\Chi_1,...,\Chi_s$.% where $\Chi_{\rho}\downarrow_H=\sum_{i=1}^r \lambda _i \Chi_i$ with $r>s$.

 Let $$B:=\sum_{g\in G}\sum_{h_1,h_2\in H}
  E(h_1)E(h_2)\Chi_{\rho}\left(h_1gh_2g^{-1}\right)$$ Then,
 
 $$B=\sum_{h_1,h_2\in H}  E(h_1)E(h_2)
 \sum_{g\in G} \Chi_{\rho}\left(h_1gh_2g^{-1}\right)$$

 Using formula (\ref{Ancien_A}) we have,

 $$B=\sum_{h_1,h_2\in H}  E(h_1)E(h_2)
 \frac{|G|\Chi_{\rho}(h_1)\Chi_{\rho}(h_2)}{\Chi_{\rho}(1)}.$$

 Since $E(h)= \sum_{i=1}^s \Chi_i(1) \conj{\Chi_i(h)}$
 and $\lambda_i$ is the multiplicity of $\Chi_i$ in the restriction of $\Chi_\rho$ to $H$ we can apply the
 orthogonality relation (\ref{orthorelations}) to get

 $$B=\frac{|G||H|^2 \left(\sum_{i=1}^s \lambda_i \Chi_i(1)\right)^2}{\Chi_{\rho}(1)}.$$

 Let $m:=\sum_{i=1}^s \lambda_i \Chi_i(1)$ and $n:=\Chi_{\rho}(1)$
 summing each side of (\ref{fonda2}) for all $g\in G$ gives  :

 $$(|G|-|H|) d_c^2=|G|m-\frac{1}{|H|^2}B$$ hence
 $$(|G|-|H|) d_c^2=|G|m-\frac{|G|m^2}{n}$$ and 
 $$d_c^2=\frac{|G|}{|G|-|H|} \left(m-\frac{m^2}{n}\right).$$
 If $N:=|G|/|H|$,  $$d_c^2=\frac{N}{N-1} \frac{m(n-m)}{n}$$
 Looking at the definitions of $m,n$ and $N$ we have exactly a packing of
 $N$ elements in $ \G_{m,n}$ which reaches the simplex bound.
\end{proof}

%-----------------
% PART V Examples
%-----------------

\section{Examples}

 This section is devoted to various examples coming from the classification of 
 2-transitive groups. We recall this classification to show that the previous construction
 gives infinite families of codes meeting the simplex bound, and to allow computation 
 of examples not handled in this paper.

\subsection{The classification of 2-transitive groups}\label{CLASS2T}

% As we have seen any group fulfilling the condition 
% $|H\backslash G/H|=2$ verifies that $G$ has
% a 2-transitive action on $G/H$. But reciprocally for any 2-transitive group
% acting on a set $\Omega$, choose $H$ to be any point stabilizers ; then
% $|H\backslash G/H|=2$. To be precise we remark that nothing 
% says that the action of $G$ on $G/H$ is faithful.
% In the case where the action is not faithful the groups $G$ (and $H$) 
% may be rewritten $G:=G_1 \ltimes N$ and $H:=H_1 \ltimes N$ where
% the action of $G_1$ on $G_1/H_1\approx G/H$ is 2-transitive and faithful.
 
 Faithful 2-transitive groups have been classified \cite{MR1721031,MR1409812} (this result
 rely on the classification of simple group). This classification (achieved in the early 80's)
 has required work of various people (Huppert, Hering, Curtis, Kantor...). To 
 summarize, there are eight types  of infinite families of 2-transitive 
 groups and some `sporadic' groups. Among the eight families four are
 quite easy to describe:
 
 \begin{enumerate}
 \item \underline{The alternating groups} : $\mathfrak{A}_n$  acting on the set 
 $\{1,\dots,n\}$ ($n-2$ transitive).
 \item \underline{The symmetric groups} : $\mathfrak{S}_n$  acting on the set 
 $\{1,\dots,n\}$ ($n$ transitive).
 \item \underline{Affine groups} : Let $V$ be the vector space $(\F_q)^d$. Affine groups
 are the groups $G:=V\rtimes G_0$ where $G_0$ is a subgroup\footnote{
 $\Gamma L_d(\F_q)$ is the group acting on $V$ generated by
 $\textrm{GL}_d(\F_q)$ and all field automorphisms,  
 $\sigma\ :\ \F_q \to \F_q$  acting component-wise on elements of $V$.} of
 $\Gamma \textrm{L}_d(\F_q)$ which verifies one of the following \mbox{conditions :}
 \begin{itemize}
 \item $\textrm{SL}_d(\F_q)\le G_0\le \Gamma L_d(\F_q)$,
 \item $\textrm{Sp}_d(\F_q)\le G_0\le \Gamma L_d(\F_q)$ where $d=2m$,
 \item  $G_0 = G_2(2^m)$.
  \end{itemize}
 There is also a finite number of special cases with dimensions $\le 6$.
 \item \underline{Projective groups} : These are the  groups $G$ with $\textrm{PSL}_d(\F_q) \le G \le
 \textrm{P}\Gamma\textrm{L}_d(\F_q)$ acting on lines of $(\F_q)^d$.
 \end{enumerate}

 There are four other families of 2-transitive groups coming from groups of Lie type.
 Describing these groups in details is not the aim of this
 article, one can use the ATLAS \cite{ATLAS} to have a permutation representation or 
 see the references for more details.

 \begin{enumerate}
 \setcounter{enumi}{4}
 \item \underline{Symplectic groups} : $G=Sp_{d}(\F_2)$  ($d=2m$) acting on subsets of
 transvections. The degree is $2^{d-1}(2^d-1)$ or $2^{d-1}(2^d+1)$.
 \item \underline{The unitary projective groups} : $\textrm{PSU}_3(q) \le G \le
 \textrm{P}\Gamma\textrm{U}_3(q)$ acting on isotypic lines of a quadratic
 form or on points of a $S(2,q+1,q^3+1)$ Steiner system.
 \item \underline{Suzuki groups} : $Sz(q)$, ($q=2^{2m+1}$), acting on points of a
 $S(3,q+1,q^2+1)$ Steiner system.
 \item \underline{Ree groups} : $R(q)$, ($q=3^{2m+1}$), acting on the points of a
 $S(2,q+1,q^3+1)$ Steiner system.
 \end{enumerate}
  There are also some `sporadic' 2-transitive groups  with
  peculiar actions. They are
  (cited with their respective degree) :
  %$(\textrm{PSL}_2(11),11),\ (\textrm{PSL}_2(8),28),$
  $(M_{11},11)$, $(M_{11},12),$ $(M_{22},22),$ $(M_{23},23),$
  $(M_{24},24),$ $(A_7,15),$ $(PSL(2,11),11),$ $(PSL(2,8),28),$
  $(HS,176),$ $(Co_3,276).$\\

  Alas the use of this classification is partially theoretical. Indeed explicit 
  representations and character tables are not known for all these 2-transitive 
  groups. But they are known for some
  infinite families (as for example $\textrm{PGL}_2(\F_q)$ or
  $\textrm{PSL}_2(\F_q)$ and their upper triangular subgroups or simply
  $\mathfrak{S}_{N-1}\subset \mathfrak{S}_N$) and then they give birth to 
  explicit infinite families of optimal simplicial configurations.
  Next subsections are devoted to present the parameters obtained with these groups.

\subsection{The symmetric group}\label{exemple1}

 Taking $G=\mathfrak{S}_N$ and $H=\textrm{Stab}(1) \cong \mathfrak{S}_{N-1}$ (so 
 $G/H \approx \{1,\hdots,N\}$), the action of $G$  is 2-transitive on $G/H$.
 The irreducible representations of $\mathfrak{S}_N$ are well known (see
 \cite{MR1824028}).

 Table \ref{tableSN} gives the parameters of codes constructed from definition \ref{CONSTRUCTION0}
 and theorem \ref{main_thm} with $G= \mathfrak{S}_N$ and  $\mathfrak{A}_N$, $4\le N \le 8$. The chordal distance is
 derived from the simplex bound (\ref{RankinSimplexBound}).
 In fact representations of the symmetric group are known well enough to give an explicit 
 and general method to obtain these parameters. With this method one can easily
 compute the parameters for any $N$. We have described this method in details in Table \ref{table_exemple}
 and have developed an example in parallel.\\
 
\begin{table*}[!]\caption{Parameters for codes coming from $\mathfrak{S}_N$ and  $\mathfrak{A}_N$, $4\le N \le 8$}\label{tableSN}
\centering
 $$ \begin{array}{|c|c|}\hline \multicolumn{2}{|c|}{N = 4} \\\hline   n & 3 \\ m &  1\\
   d_c^2 & \frac{8}{9} \\\hline \end{array} \quad 
 \begin{array}{|c|c|c|c|}\hline \multicolumn{4}{|c|}{N = 5} \\\hline   n & 4 & 5 & 6 \\ m &  1 & 1,2 & 3 \\
   d_c^2 & \frac{15}{16} & 1 ,  \frac{3}{2} & \frac{15}{8}\\\hline
\end{array}
 \quad
 \begin{array}{|c|c|c|c|c|c|}\hline \multicolumn{6}{|c|}{N = 6}\\\hline   n & 5 & 8 & 9 & 10 & 16 \\ m &  1 & 3 & 4 & 4 & 5,6 \\
 d_c^2 & \frac{24}{25} & \frac{9}{4}  & \frac{8}{3} & \frac{72}{25} & \frac{33}{8}  , \frac{27}{4} \\\hline     
\end{array}$$

$$\begin{array}{|c|c|c|c|c|c|c|}\hline \multicolumn{7}{|c|}{N = 7} \\\hline   n & 6 & 14 & 15 & 20 & 21 & 35 \\ m &  1 & 5 & 5 & 10 & 5,8 & 8,9,10,16,17 \\ d_c^2 & \frac{35}{36} & \frac{15}{4} & \frac{35}{9} & \frac{35}{6} & \frac{40}{9} , \frac{52}{9} & \frac{35}{6},\frac{39}{5},\frac{25}{3},\frac{152}{15},\frac{51}{5}   \\\hline   \end{array}$$

$$\begin{array}{|c|c|c|c|c|c|c|c|c|c|c|c|}\hline  
\multicolumn{12}{|c|}{N = 8} \\\hline   n & 7 & 20 & 21 & 28 & 35 & 42 & 45 & 56 & 64 & 70 & 90  \\ m &  1 & 6 & 6 & 14 & 10,15 & 21 & 10 & 21 & 14,15,35 & 14,21,35 & 20,35\\d_c^2 & 
\frac{45}{49} & \frac{24}{5} & \frac{240}{49} & 8 & \frac{400}{49},\frac{480}{49} & 12 & \frac{80}{9} & 15 & \frac{25}{2},
\frac{105}{8},\frac{145}{8} & \frac{64}{5} , \frac{84}{5} , 20 & \frac{160}{9}, \frac{220}{9}
 \\\hline \end{array}
$$
\end{table*}

\begin{table*}[!]\caption{The example of $\mathfrak{S}_N$}\label{table_exemple}\centering 
 \begin{tabular}{|c|c|}
  \hline
  Facts & Example \\  &\\
\begin{tabular}{p{8cm}}$\bullet$ There is an irreducible representation of
 $\mathfrak{S}_N$  for each partition $\lambda=[\lambda_1,\dots,\lambda_k]$
 of n (\ie{} a decreasing sequence of integers $[\lambda_1,\dots,\lambda_k]$
 whose sum is $N$). \end{tabular}
&
\begin{tabular}{c} N:=12\\ \\
$\lambda:=[6,4,2]$\end{tabular}\\
&\\
\begin{tabular}{p{8cm}}$\bullet$ One can associate a diagram to a partition
 in the following way. Draw $\lambda_1$ box on the first line, $\lambda_2$
 on the second... As in the example. \end{tabular}
&
\begin{tabular}{c}
 \includegraphics[height=1cm]{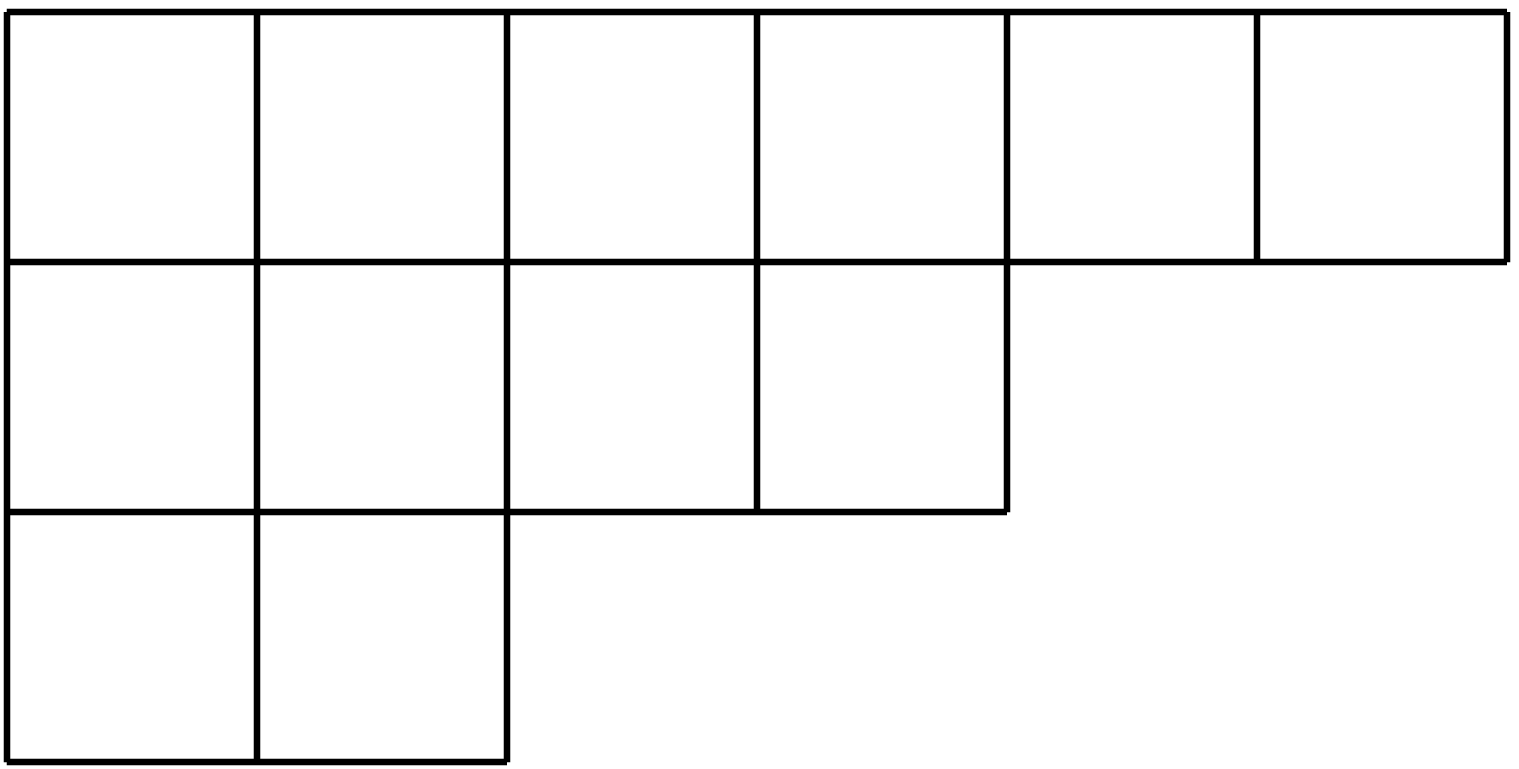}
\end{tabular}\\
&\\

\begin{tabular}{p{8cm}}$\bullet$ The hook length of a box is the sum of the
 number of boxes under it (in the same column) and at its right (in the same
 line) plus one. In the example, we have filled each box with the length of
 the associated hook.
\end{tabular}
&
\begin{tabular}{c}
 \includegraphics[height=1.6cm]{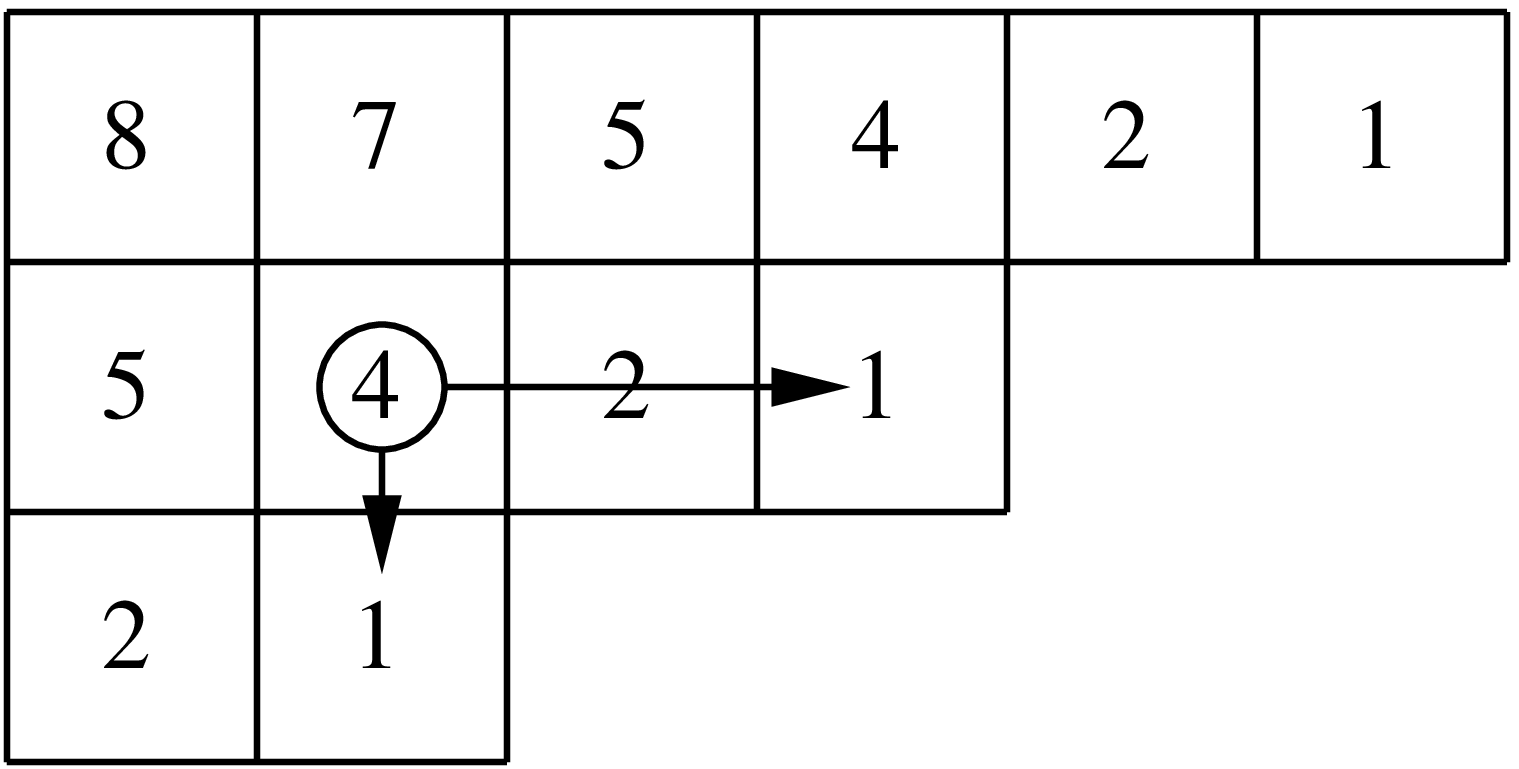}
\end{tabular}\\

&\\
\begin{tabular}{p{8cm}}$\bullet$ The dimension of the representation
 associated to $\lambda$ is given by $\frac{N!}{z}$ where $z$ is the product
 of the hook length of every box. \end{tabular}
&
\begin{tabular}{c}
 $\dim(\Chi_\lambda)=2673$
\end{tabular}\\

&\\
\begin{tabular}{p{8cm}}$\bullet$ The branching rule (see \cite{MR1824028})
 states that when restricted to $\mathfrak{S}_{N-1}$ the irreducible
 character associated to lambda decomposes itself as
 $\Chi_{\lambda}=\Chi_{\mu_{(1)}}+\dots+\Chi_{\mu_{(\ell)}}$ where each
 $\mu_{(i)}$ is  obtained from $\lambda$ by deleting a `corner' box.
\end{tabular}
&
\begin{tabular}{c}
 \includegraphics[height=1.4cm]{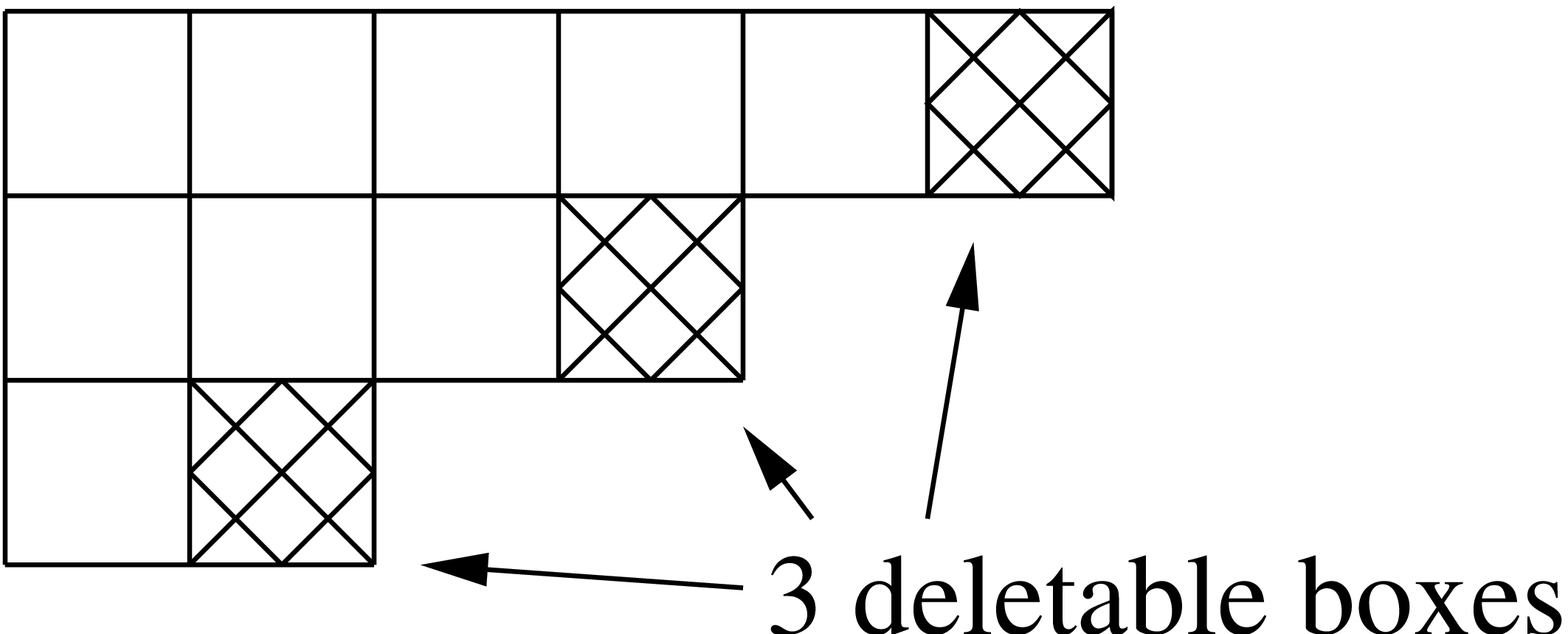}\\
 $\Chi_{[6,4,2]}\downarrow^{\mathfrak{S}_N}_{\mathfrak{S}_{N-1}} =$\\
 $\Chi_{[5,4,2]}+\Chi_{[6,3,2]}+\Chi_{[6,4,1]}$
\end{tabular}\\
&\\
\begin{tabular}{p{8cm}}$\bullet$ Now we have everything needed to compute
 the parameters obtained. Let us choose the isotypic space associated to
 $\Chi_{[5,4,2]}$. Its dimension can be computed with the hook formula and
 have a configuration with the following parameters.
\end{tabular}
&
\begin{tabular}{c}
 $N=12, \ n=2673, \ m= 990$\\
 so $d_c^2=680$\\ (reaching the Simplex bound)
\end{tabular}\\
\hline
\end{tabular}
\end{table*}

\subsection{The groups $\textrm{PGL}_2(\F_q)$ and $\textrm{PSL}_2(\F_q)$} 
 We have fully studied all the constructions 
 obtained with $\textrm{PGL}_2(\F_q)$ and $\textrm{PSL}_2(\F_q)$ ($q$ odd). In this case the subgroup $H$ 
 can be chosen as the subgroup of upper triangular matrices. Explicit descriptions of irreducible 
 representations and characters of these groups can be found in \cite{MR1153249,MR1864147,MR793377,MR696772}.
 So we can give all the simplicial configurations coming from representations and characters of these groups.
 For all $q$ such that $\F_q$ exists (\ie{} $q$ is a power of a prime) we have Grassmannian
 configurations with parameters as given in Table \ref{tableGLFQ}.

 \begin{table*}[!]
 
\caption{Explicit parameters coming from $\textrm{PGL}_2(\F_q)$ and $\textrm{PSL}_2(\F_q)$ ($q$ odd)}\label{tableGLFQ}
\centering
{ $N:=q+1$ }\\
\begin{tabular}{|c|c|c|c|c|}
\hline
$n$ & $q+1$ & $q$ & $q+1$ & $q$ \\
\hline
$m$ & $1$ & $1$ & $2$ & $\frac{q-1}2$\\
\hline
$d_c^2$ & $1$ & $1-\frac{1}{q^2}$ & $2\frac{q-1}{q}$ & $\frac{(q+1)^2(q-1)}{4q^2}$ \\
\hline
$\tilde{d}$ & $1$ & $1-\frac{1}{q^2}$ & $(\frac{q-1}{q})^2$
& $?$ \\
\hline
\end{tabular}

\vspace{10pt}

\begin{tabular}{|c|c|c|c|c|}
\hline
$n$ & $q+1$ & $q+1$ & $\frac{q+1}2$ & $q-1$\\
\hline
$m$ & $\frac{q-1}{2}$ & $\frac{q+1}{2}$ & $1$ & $\frac{q-1}2$\\
\hline
$d_c^2$ & $\scriptstyle{\frac{(q-1).(q+3)}{4.q}}$ & $\scriptstyle{\frac{(q+1)^2}{4.q}}$ & $\frac{q-1}{q}$ &$\scriptstyle{\frac{q^2-1}{4.q}}$\\
\hline
$\tilde{d}$ & $?$ & $?$ & $\frac{q-1}{q}$ & $?$\\
\hline
\end{tabular}
\end{table*}

 We give the value of $\tilde{d}$ when it
 is known. This value is difficult to evaluate in the cases where a "?" appear. Indeed numerical experiments show
 that depending on $q$ this value can be 0. We were unable to make any conjecture about its behavior. For example the sets of principal angles corresponding to the last column configuration are given in Table \ref{GL_ANGLES} 
 (we give the values of $\sin^2(\theta_i)$)

 \begin{table*}[!]
\caption{Explicit values of principal angles}\label{GL_ANGLES}
$$\begin{array} {|c|c|} 
 \hline
\multicolumn{2}{|c|}{\substack{n=q-1,\ m=\frac{q-1}{2}\\N=q+1,\ d_c^2=\frac{q^2-1}{4q} }}\\
 \hline
 q=5 & ( 1, \frac{1}{5} )\\
 \hline
 q=7 & (0,\frac{6}{7},\frac{6}{7})\\
 \hline
 q=9 & (1,\frac{5}{9},\frac{1}{3},\frac{1}{3})\\
 \hline
 q=11 & (0,\underbrace{\frac{7+3\sqrt{5}}{22} , \frac{7-3\sqrt{5}}{22}}_{\textrm{twice}}) \\
 \hline
 \end{array}$$
 \end{table*}

\subsection{Some Symplectic and Sporadic groups}

 In the examples of previous sections, representations and characters are well known. So an explicit description
 of the codes (by projection matrices for instance) is possible and the values of principal angles can be given.
 This may not be the case for codes coming from high degree irreducible representations of big symplectic 
 groups or simple groups (such as $Sp_{10}(2)$ or $Co_3$).
 Nevertheless one may be interested in small parameters coming from such groups given in Tables 
\ref{tableSYMPLECT} and \ref{tableSIMPLE}. The chordal distance is recorded when its rational form involves
 coefficients with few digits. In any case it can be derived from the simplex bound (\ref{RankinSimplexBound}).\\

 To compute these tables we used permutations to describe these groups. The degree of the permutation 
 representations\footnote{These descriptions of the Symplectic and Sporadic groups can be found in the ATLAS \cite{ATLAS}.}
 was chosen according to the classification of 2-transitive groups (see section \ref{CLASS2T}). Then the subgroup $H$ 
 can be chosen as $\textrm{Stab}(1,G)$.\\

\begin{table*}[!]
\caption{Small parameters coming from small symplectic groups}\label{tableSYMPLECT}\centering
\begin{small}

$$\begin{array}{|c|c|c|c|c|c|c|c|}
\hline 
\multicolumn{5}{|c|}{Sp_4(2) \quad N=10 }\\
\hline 
n & 5 & 8 & 9 & 10 \\
m & 1& 4 & 1,4 & 1,2,4,5 \\
d_c^2 & \frac{8}{9} &  \frac{20}{9} & \frac{80}{81},\frac{320}{81} & 1,\frac{16}{9},\frac{8}{3},\frac{25}{9}\\
\hline 
\end{array}$$

$$\begin{array}{|c|c|c|c|c|c|c|c|}
\hline 
\multicolumn{5}{|c|}{Sp_4(2) \quad N=6  }\\
\hline 
n & 5 & 8 & 9 & 10 \\
m & 1& 3 & 4 & 3,4 \\
d_c^2 & \frac{24}{25} & \frac{9}{4} & \frac{8}{3} & \frac{63}{25} , \frac{72}{25} \\
\hline 
\end{array}$$

$$\begin{array}{|c|c|c|c|c|c|c|c|}
\hline 
\multicolumn{8}{|c|}{Sp_6(2) \quad N=36 \quad n\le 100 }\\
\hline 
n &  15 & 21 & 27 & 35 & 56 & 70 & 84 \\
m & 1 & 1 & 7 & 1,14, 15 & 7,21,28 & 28& 14, 28 \\
d_c^2 & \frac{24}{25} & \frac{48}{49} & \frac{16}{3} & \frac{1224}{1225},\frac{216}{25},\frac{432}{49} & 
 \frac{63}{10},\frac{27}{2},\frac{72}{5} & \frac{432}{25} & 12 , \frac{96}{5}\\
\hline 
\end{array}$$

$$\begin{array}{|c|c|c|c|c|c|c|c|}
\hline 
\multicolumn{8}{|c|}{Sp_6(2) \quad N=28 \quad n\le 100 }\\
\hline 
n & 7 & 21 & 27 & 35 & 56 & 70 & 84 \\
m & 1 & 1,6 & 1,6,7 & 15 &  6,20,26 & 10 & 24 \\ 
d_c^2 & \frac{8}{9} & \frac{80}{81},\frac{40}{9} & \frac{728}{729},\frac{392}{81},\frac{3920}{729} & \frac{80}{9} & 
\frac{50}{9},\frac{40}{9},\frac{130}{9} & \frac{8}{9} & \frac{160}{9}\\
\hline 
\end{array}$$

$$\begin{array}{|c|c|c|c|c|c|c|c|c|}
\hline 
\multicolumn{9}{|c|}{Sp_8(2) \quad N=136 \quad n\le 1000 }\\
\hline 
n & 51 & 85 & 119 & 135 & 238 & 510 & 595 & 918 \\
m & 1 & 1 & 35 & 1,50,51 & 28 & 210 & 28,175 & 50,168,218  \\ 
d_c^2 & \frac{80}{81} & \frac{224}{225} & \frac{224}{9} & ... & \frac{224}{9} & \frac{1120}{9} & \frac{627}{25},\frac{1120}{9} & ... \\
\hline 
\end{array}$$

$$\begin{array}{|c|c|c|c|c|c|c|c|c|}
\hline 
\multicolumn{9}{|c|}{Sp_8(2) \quad N=120 \quad n\le 1000 }\\
\hline 
n & 35& 85 & 119 & 135 & 238 & 510 & 595 & 918 \\
m & 1 & 1 & 1,34,35 & 51 & 34 & 34 & 34,204,238 & 204 \\ 
d_c^2 & \frac{48}{49} & \frac{288}{289} & \frac{944}{945},\frac{4624}{189},\frac{224}{9} & ... & \frac{9248}{315} & ... & ...& ...\\
\hline 
\end{array}$$

$$\begin{array}{|c|c|c|c|c|c|}
\hline 
\multicolumn{6}{|c|}{Sp_{10}(2) \quad N=528 \quad n\le 10000 }\\
\hline 
n & 187 & 341 & 495 & 527 & 6138 \\
m & 1 & 1 & 155 & 1,186,187 & 868 \\ 
d_c^2 & \frac{288}{289} & \frac{960}{961} & \frac{320}{3} & ... & \frac{2240}{3}\\
\hline 
\end{array}$$

$$\begin{array}{|c|c|c|c|c|c|}
\hline 
\multicolumn{6}{|c|}{Sp_{10}(2) \quad N=496 \quad n\le 10000 }\\
\hline 
n & 155 & 341 & 495 & 527 & 6138 \\
m & 1 & 1 & 1,154,155 & 187 & 154,748,902 \\ 
d_c^2 & \frac{224}{225} & \frac{1088}{1089} & ... & \frac{1088}{9} & ... \\
\hline 
\end{array}$$
\end{small}
\end{table*}

\begin{table*}[!]
\caption{Small parameters coming from some sporadic groups}\label{tableSIMPLE}\centering
\begin{small}
 $$\begin{array}{|c|c|c|c|c|c|c|c|c|c|}
\hline 
\multicolumn{10}{|c|}{Co_3 \quad N=276}\\
\hline  n & 23 & 253 & 275 & 1771 & 2024 & 4025 & 5544 & 7084 & ...\\\hline m &  1 & 1,22 & 1,22,23 & 231 & 22,252,274 & \substack{\\22,231,252,253,\\274,483,505} &  
\substack{\\22,252,274,1750,\\1772,2002,2024} & 1540 &... \\
d_c^2 & \frac{}{}\frac{24}{25} & \frac{3024}{3025},\frac{504}{25} & ... & \frac{1008}{5} & ... & ... & ... & \frac{6048}{5}& ...\\
\hline \end{array}
$$

$$\begin{array}{|c|c|c|c|c|c|c|c|c|}\hline 
\multicolumn{8}{|c|}{HS \quad N=176}\\
\hline
  n & 22 & 77 & 154 & 175 & 231 & ... & 3200 \\\hline m &  1 & 21 & 1,21,28,29,49 & 1,21,22,28,29,49,50 & 21,84,105 & ... & \substack{\textrm{more than } 200\\ \textrm{possibilities}} \\
d_c^2 & \frac{24}{25} & \frac{384}{25} & \frac{1224}{1225},\frac{456}{25},\frac{576}{25},\frac{1160}{49},\frac{168}{5} & ... & 
\frac{96}{5},\frac{1344}{25},\frac{288}{5} & ... & ...\\
\hline \end{array}
$$

$$\begin{array}{|c|c|c|c|c|c|c|c|c|c|}\hline 
\multicolumn{9}{|c|}{M_{24} \quad N=24}\\
\hline
  n & 23 & 252 & 483 & 1035 & 1265 & 1771 & 2277 & ... \\\hline m &  1 & 22 & 230 & 45 & 230 & 231,770 & 253 & ...\\
d_c^2 & \frac{528}{529} & \frac{440}{21} & \frac{880}{7} & ... & \frac{2160}{11} &... & \frac{704}{3} & ...\\
\hline \end{array}
$$
\end{small}
\end{table*}

%-------------------
% PART VI remarks
%-------------------

\section{Generalizations}

 In this section we discuss various generalizations of the previous result.
 First we give a method to expand the set of codes reaching the simplex bound.
 Then we have a look at unions of configurations coming from group orbits.
 Finally we give a new insight on some optimal orthoplex configurations given in \cite{AGroup}.

 \subsection{How to extend the set of parameters}\label{pareil}

 We have already found a large set of parameters for codes reaching the simplex bound
 but it may be interesting to extend it further.
 Thinking in terms of projection matrices one can have the idea to use
 Kroenecker product ($\otimes$). As a first try we can make the product of
 all matrices of a configuration with the identity matrix of rank $k$ :
 $I_k$. $$\left\{ I_k\otimes \Pi_{g.W} \ :\ g \in G/H \right\}$$
 One can easily see that this multiplies $m,n,d_c^2$ by $k$ and keeps $N$
 invariant. So we have the following :

 \begin{prop}
 If we have an explicit configuration with parameter $N,m,n$ which reach the simplex bound,
 then for any positive integer $k$ it is possible to build a new 
 optimal configuration with parameters $N,k.m,k.n$.\\
 \end{prop}

 This idea can be extended to couple of configuration. Let $\{ \Pi_{1,i} : i \in [\![ 1,\dots,N_1]\!] \}$ and 
 $\{ \Pi_{2,j} : j \in [\![ 1,\dots,N_2]\!] \}$ be two configurations in
 $G_{m_1,n_1}$ and $G_{m_2,n_2}$ with squared minimal distances equal to $d_{c1}^2$ and $d_{c2}^2$. The eigenvalues of $(\Pi_1\otimes \Pi_2)(\Pi'_1\otimes \Pi'_2)$ are the products
 $\lambda_i\mu_j$ where $\lambda_i$ and $\mu_j$ are the eigenvalues of $\Pi_1\Pi'_1$ and $\Pi_2 \Pi'_2$ respectively.
 One can then deduce that the new chordal distance is $\min(m_1d_{c2}^2,m_2d_{c1}^2)$.

 \subsection{Minimal distance in unions}

 In this section we study in unions of configurations obtained 
 by orbits of isotypic subspaces. We first state a variation of proposition \ref{EQANGLES} (with a similar proof).

 \begin{prop}\label{EQANGLES2}
 Let $W_1$ and $W_2$ be two subspaces of $\C^n$ stable under the action of $H$.
  The number of different sets of principal angles between $g_1W_1$ and $g_2W_2$ ($g_1,\ g_2 \in G$) is
  bounded above by the number of orbits for the action of  $G$ on ordered pairs of $G/H$. Moreover the set of principal angles 
 for the pairs $(W_1,g_0W_2)$ and $(W_1,gW_2)$  is the same for all $g\in Hg_0H$. \\
 \end{prop}

 This property can be used to find the minimal chordal distance in unions
 of orbits. We can for example use it with configurations coming from 2-transitive groups, this gives the following theorem.

 \begin{thm}
  Let $G$ be a group and $H$ a subgroup such that $G$ act 2-transitively by left
 multiplication on $G/H$. Suppose furthermore that we dispose of an
 irreducible representation  
 $\rho : G \to \textrm{GL}_n(\C)$ which is reducible when restricted to
 $H$. If $W_1,...,W_t$ are direct sums of $H$-isotypic subspaces associated to disjoints subsets of 
 irreducible characters, all $W_i$'s having the same dimension $m$, then 
 the orbit of $W_1,...,W_t$ under $G$ gives a code of cardinal $N=t|G/H|$ having minimal distance
 $d_c^2= \frac{N}{N-1}\frac{m(n-m-n/N)}{n}$.
 \end{thm} 

 \begin{proof}
 Suppose that
 $\Chi_{\rho}\downarrow^G_H=\sum_{i=1}^s \lambda_i.\Chi_i$, and that we have found disjoints subsets
 $S_1,\dots,S_t$ of $ [\![ 1,\dots,s]\!] $ such that
 $$ \forall j\in [\![ 1,\dots,t]\!],\  \sum_{i\in S_j} \lambda_i.\Chi_i(1)=m$$ for a fixed $m$.

 We need to compute the distance value for any pair of $m$
 dimensional planes. Let $W_1$ and $W_2$ be the direct
 sums of isotopic subspaces associated to $S_1$ and $S_2$ and focus on 
 the distance between $W_1$ and $g.W_2$.\\

 If $g\in H$ then $g.W_2=W_2$ and the two spaces $W_1$ and $W_2$ are orthogonal so 
 the minimal distance between them is $m$. If $g\notin H$ the chordal distance may take only
 one value by Proposition \ref{EQANGLES2}.
 
 If we sum over $g\in G$ the analog of (\ref{fonda2}) with different characters 
 we get the following equality :
 $$|H|m+(|G|-|H|)d_c^2(W_1,gW_2)=|G|m-\frac{1}{|H|^2}B$$
 where 
 $$B=\sum_{g\in G\ } \  \sum_{\ h_1,h_2\in H} E_{S_1}(h_1)E_{S_2}(h_2)\Chi_{\rho}(h_1gh_2g^{-1})$$
and
 $$
 E_{S_j}(h):=\left(\sum_{i \in S_j}{\displaystyle \Chi}_i(1)\conj{{\displaystyle  \Chi}_i(h)}\right)
 $$ 
 Following the proof of Theorem \ref{main_thm}, we found that
 $$B=\frac{\displaystyle |G||H|^2\left(\sum_{i \in S_1} \lambda_i \Chi_i(1)\right)\left(\sum_{i \in S_2} \lambda_i \Chi_i(1)\right)}{\Chi_G(1)}$$

 so
 $$d_c^2(W_1,gW_2)= \frac{N}{N-1}\frac{m(n-m-\frac{n}{N})}{n}.$$ This value
 is smaller than $d_c^2(W_1,W_2)=m$ or $d_c^2(W_1,gW_1)$.
 It is therefore the minimal chordal distance.\\
 \end{proof}

 We may carry on the example of Table \ref{tableSN}. One can observe that the two
 components $\Chi_{[5,4,2]}$ and $\Chi_{[6,3,2]}$ in the decomposition of
 $\Chi_{[6,4,2]}\downarrow^{\mathfrak{S}_{12}}_{\mathfrak{S}_{11}}$ have the
 same dimension (990). So if we take the union of the two configurations we
 get a configuration of $N=24$ planes of dimension $m=990$ in $\C^{2673}$
 with minimal distance $d_c^2=\frac{13970}{23}\approx 607.4$ (for a value
 of 650.4 of the simplex bound). This may lead to not so bad configurations
 especially if we have a lot of subspaces with the same dimension (for the
 new value of minimal distance does not depend on how many sets are 
 joined).\\

 \subsection{An optimal orthoplex configuration}
  In this section we show how a small variation
  of the method described above gives back the optimal configurations presented
  in \cite{AGroup} (Thm 1). We first recall the construction of
  the keystone groups :

   Let $U=\F_2^i$ and $V=\R^n$ where $n=2^i$. Let $\{ e_u :u \in U\}$ be a 
  vector basis for $V$ and let $E$ be the (extraspecial) subgroup of the
  orthogonal group $\mathcal{O}=O(V)$ generated by 
  \begin{small}
  $$X(a) : e_u \mapsto e_{u+a},\ \textrm{ and } Y(b) : e_u \mapsto (-1)^{b.u} e_u, \quad u \in U.$$
  \end{small}
  The normalizer $L$ of $E$ in $\mathcal{O}$ is the (Clifford type) subgroup
  of $\mathcal{O}$ generated by\footnote{See \cite{MR1455862}.}:
  $E,\ H,\ \tilde{H}_2,\ \widetilde{GL (V)} \textrm{ and } \{d_M : M \textrm{ skew-symmetric}\}$,
  where 
  \begin{itemize}
   \item $H=\frac{1}{\sqrt{N}}[(-1)^{u.v} ]_{u,v \in V}$
   \item $\tilde{H}_2=\frac{1}{\sqrt{2}}\begin{scriptsize}\left(\begin{array}{cc} 1 & 1 \\ 1 & -1\end{array}\right)\end{scriptsize} \otimes I_{2^{(i-1)}}$ 
   \item $\widetilde{GL(V)}$ is the group generated by the orthogonal transformations $\{G_A : A \in \textrm{GL}(U)\}$
  where $G_A : V \to V$ permute the coordinates by sending $e_u$ on $e_{Au}$.
   \item $d_M$ is the diagonal matrix $(-1)^{Q_M(v)}$ where $Q_M$ is the quadratic form associated to the skew-symmetric matrix 
       $M$ i.e. $Q_M(u+v)=Q_M(u)+Q_M(v)+uMv^T$.
  \end{itemize}

 Note that this description of the group $E$ is a unitary representation and its character $\Chi_E$
 as value 0 for any element except $\pm I$.

 Let $\mathcal{S}_r$ be the set of abelian subgroups in $E$ generated by $-I$ and $r$ independent order 2 
 element $g_1,...,g_r$ of $E$ and $S \in \mathcal{S}_r$.
 Then the restriction of $\Chi_E$ to $S$ is equal to 
 the sum of $2^r$ distinct linear characters with multiplicity $2^{i-r}$.\\

 The configuration described in \cite{AGroup} (Thm 1) is the set of all isotypic subspaces
 for all subgroups $S$ in $\mathcal{S}_r$ and all their characters.\\

 Fix $S\in \mathcal{S}_r$ and $\chi$ an irreducible character and let 
 $\Pi_{S,\Chi}=\frac{1}{|S|}\sum_{s\in S} \conj{\Chi(s)}s$ 
 be the projection on the associated isotypic subspace $W$. Considering the orbit of $W$ under $L$ (instead of $E$),
 one can check that 
 $$\left\{\rho(g)\Pi_{S,\Chi}\rho(g^{-1})=\frac{1}{|S|}\sum_{s\in gSg^{-1}} \conj{\Chi(g^{-1}sg)}s\ \ : \ \ g\in L \right\}$$
 is exactly the set of projection matrices of the the Grassmannian code described in \cite{AGroup} (Thm 1).
 This is a consequence of the following facts : \\
 - the action of $L$ by conjugation on $\mathcal{S}_r$ is transitive ; \\
 - the action of $E$ on characters of $S$ given by $g.\chi(x)= \chi(gxg^{-1})$ is transitive.\\

 Let now focus on the optimal case where $r=1$ (i.e. $|S|=4$), the optimality of this configuration 
 is a special case of the following proposition :
 \begin{prop}
 Let $G$ be a subgroup of $U_n(\C)$ such that the only elements having a non-zero trace are $\pm I$ . 
 Let $\mathcal{S}$ be the set of distinct abelian subgroups generated by $-I$ and another element  $g$ of order $2$.
 Then the set of isotypic subspaces with projection matrices 
 $$\frac{1}{|S|}\sum_{s\in S} \Chi_1(s)s \quad \textrm{ and } \quad 
 \frac{1}{|S|}\sum_{s\in S} \Chi_2(s)s \quad \textrm{ for } S \in \mathcal{S}$$
 where $$\Chi_i(-1)=-1 \quad \textrm{and} \quad \Chi_i(g)=(-1)^i$$
 form a Grassmannian code in $\G_{m,n}$ with $m=\frac{n}{2}$ and cardinality $2|\mathcal{S}|$ in which the 
 only non-zero distances are $m$ and $m/2$. This code reaches the orthoplex bound when $2|\mathcal{S}| > \frac{n(n+1)}{2}$.
 \end{prop}

 \begin{proof} For convenience we suppose that the above group is obtained by a representation
 $\rho : G \to U_n(\C)$.
 We first remark that $W=\frac{1}{|S|}\sum_{s\in S} \Chi_i(s)s$ is an isotypic subspace.
 The restriction of the matrix representation to $S$ is reducible. It is clear that
 its character decomposition is  $$\Chi_\rho \downarrow^G_S =\frac{n}{2} \Chi_1 + \frac{n}{2} \Chi_2.$$
 So the isotypic subspaces have dimension $m=\frac{n}{2}$ (and $n$ is divisible by 2).
 Let $W_1:=\frac{1}{|S_1|}\sum_{s\in S_1} \Chi(s)s$ and $W_2:=\frac{1}{|S_2|}\sum_{s\in S_2} \Chi'(s)s$
 be two subspaces defined by the subgroups $S_1$ and $S_2$ in $\mathcal{S}$, and the irreducible
 characters $\Chi$ and $\Chi'$. If the subgroups are equal and the characters distinct then the two isotypic
 subspaces are orthogonal and the chordal distance between $W_1$ and $W_2$ is equal to $m$. Otherwise
 $$d_c^2(W_1,W_2)= m - \frac{1}{|S_1||S_2|} \sum_{\substack{s_1\in S_1\\s_2\in S_2}}\Chi(s_1)\Chi'(s_2) \Chi_\rho(s_1s_2).$$
 Since  $\Chi_\rho(s_1s_2)$ is not zero if and only if  $s_1$ and $s_2$ are in $\{\pm I\}$ we have.
 $$d_c^2(W_1,W_2)= m - \frac{1}{16} \sum_{s_1,s_2 \in \{\pm I\}}\Chi(s_1)\Chi'(s_2) \Chi_\rho(s_1s_2).$$
  Hence
 $$d_c^2(W_1,W_2)= m - \frac{4}{16} n = \frac{m}{2}.$$
 Since the expression of the orthoplex bound (valid only if $N>\frac{n(n+1)}{2}$) is  
 \begin{equation}\label{OBOUND}
 d_c^2\le \frac{m(n-m)}{n}
 \end{equation}
 it is clear that the  above construction gives an optimal Grassmannian configuration as soon as 
 $2|\mathcal{S}| > \frac{n(n+1)}{2}$.
 \end{proof}

%---------------------
% PART VII Conclusion
%---------------------

\section{Conclusion}

 In this article we have studied some interesting Grassmannian packings obtained as group orbits.
 %Noticing the link between properties of the stabilizer and the number
 %of set of  principal angles 
 We proved that orbits of isotypic subspaces
 associated to maximal subgroups of 2-transitive groups are optimal Grassmannian configurations w.r.t. the chordal distance.
 Indeed these configurations reach the simplex bound.
 Based on the classification of 2-transitive groups we have illustrated this result with many examples
 - among which some infinite families - 
 for which we have computed the parameters.
% We illustrated this result on many examples coming from the classification 
% of 2-transitive groups, giving infinite families of optimal explicit configurations,
%  some theoretical values of parameters, and generalizations.
 %% In this article we have presented an original point of view to 
 %% construct Grassmannian codes as group orbits. Indeed, we have not 
 %% searched directly for a good starting element for the orbit.
 %% Instead we have focused on finding interesting stabilizers and have deduced
 %% interesting starting elements. This approach 
 %% has been successful in finding optimal simplicial configurations w.r.t. 
 %% the chordal distance but also to recover optimal configurations of \cite{AGroup}
 %% with a slight generalization. 
 We have also recovered optimal configurations of \cite{AGroup} in a more general context.
 If our configurations perform well 
 regarding the chordal distance they have a less obvious behavior regarding the
 pseudo-distance $\tilde{d}$. According to \cite{MR2245117} theses codes are  
 accurate for low rate communications over extremely noisy channels.\\

\section*{Acknowledgment}
The author would like to thank Christine Bachoc for her precious advises on the  writing of this article.

\end{document}